\documentclass[11pt]{article}
\usepackage{graphicx}
\title{\bf New applications of the renormalization group method in physics -- a brief introduction}
\author{Y. Meurice$^*$, R. Perry$^\dagger$, and S. -W. Tsai$^\P$\\ 
$^*$Department of Physics and Astronomy, The University of Iowa\\
Iowa City, Iowa 52242, USA \\
$^\dagger$Department of Physics, The Ohio State University\\ Columbus, OH 43210, USA\\
$^\P$Department of Physics and Astronomy, University of California\\ Riverside, CA 92521, USA}
\date{\today}
\begin{document}
\maketitle 
\begin{abstract}
The renormalization group method developed by Ken Wilson more than four decades ago has revolutionized the way we think about 
problems involving a broad range of energy scales such as phase transitions, turbulence, continuum limits and bifurcations in dynamical systems.
The theme issue provides articles reviewing recent progress made using the renormalization group method in atomic, condensed matter, nuclear and particle physics. 
In the following we introduce these articles in a way that emphasizes common themes and the universal aspects of the method. 
\end{abstract}
\begin{center}
{\bf keywords: renormalization group, phase transitions, critical phenomena, conformal symmetry, critical exponents, effective interactions}
\end{center}
\section{Introduction}
Key concepts of contemporary theoretical physics such as universality, self-similarity, scaling, data collapse and 
asymptotic freedom are directly associated with the development of the renormalization group (RG) ideas \cite{wilson74,RevModPhys.46.597,wilson83}. 
The RG method consists in ``thinning down'' the degrees of freedom in problems involving a broad range of energy scales which cannot be separated easily. 
In short, one tries to replace the effects of rapidly oscillating modes by a modification of the interactions among modes oscillating less rapidly. 
The fixed points of this procedure have universal properties that can be observed in systems with very different microscopic dynamics. For instance, 
the magnetization of iron and the vapor-liquid transition in water follow very similar scaling laws near the end of the line of first order phase transition \cite{RevModPhys.39.395}. For pedagogical introductions, we recommend the textbooks of 
Goldenfeld \cite{Goldenfeld:1992qy} and Cardy \cite{Cardy:1996xt}. 

The RG provides a universal language spoken by scientists working in very different areas.  
Despite this universality,  applications have often been developed independently  without much communication among them. 
Generating closer interaction among RG practitioners who 
are working in atomic, condensed matter, nuclear and particle physics  
was the goal of a recent  workshop held at the Institute for Nuclear Theory located in Seattle \cite{int}. 

The workshop was  organized around themes of common 
interest such as effective interactions, universality and conformal symmetry and  
succeeded to bring  communication of new exciting results across traditional boundaries.
The workshop ran over five days with 35 participants. 
Several speakers were asked to provide pedagogical introductions to 
specific areas so that people from other fields had the necessary 
background for the more technical talks. 
The goal of this theme issue is to recreate 
the event for a wider audience and provide a
snapshot of the diversity of applications of the method. 
In the following, we introduce successively articles regarding recent progress in condensed matter, nuclear and particle physics and then present various applications 
of the so-called exact RG equations. 

\section{Applications in condensed matter physics}

In quantum condensed matter systems, the different types of two-body interactions that are present are often known, but the emerging many-body physics, -- Fermi liquid behavior, superconductivity, magnetic order, charge order, to mention a few conventional examples, -- can be very rich.  Given a microscopic Hamiltonian, a full solution of the quantum many-body problem at all energy scales is in general not possible. The RG method is used to obtain effective theories at low energies. R. Shankar \cite{rs} gives an introduction to the RG for interacting non-relativistic fermions which is based on elimination of high-energy-modes around the Fermi surface, and the effect of this on the effective action. For a circular (spherical) Fermi surface in 2d (3d), the marginal interactions are of only two kinds, corresponding to the BCS and the forward channels, and in this case the RG flow equations for the coupling functions simplify considerably. The forward couplings do not flow at one-loop. The flow equation of the BCS couplings $V_m$, where the angular momentum channel is denoted by $m$, takes the generic form (in suitable units): 
\begin{equation}
\frac{dV_m}{dt} = - V_m^2 \ .
\end{equation}
and can be solved analytically, giving $V(t) = V(0)/[1 + t V(0)]$, with $t=\ln[\Lambda_0/\Lambda]$ and $\Lambda$ is the energy cut-off which decreases as the high energy modes are eliminated. For a repulsive interaction ($V(0)>0$), the effective interaction $V(t)$ is reduced when only the modes near the Fermi surface are kept, flowing logarithmically to zero, and Landau's Fermi liquid emerges as the fixed point of the RG flow. If $V(0)$ is attractive, the Fermi liquid state is no longer the ground state of the system: the analytical solution for $V(t)$ becomes singular at $t_c = 1/|V(0)|$ signaling a BCS pairing instability. This singularity is what particle physicist would call a Landau pole. The system now flows into a new fixed point corresponding to a superconductor. The value of $t_c$ sets a critical energy scale $\Lambda_c \sim \exp(-1/|V(0)|)$, with the exponential dependence as expected from the BCS theory. 
%The full form of $\Lambda_c$, with the BCS dependence at weak-couplings, McMillan's expression at weak to intermediate couplings, and Allen-Dynes equation for strong couplings can be obtained by an RG approach \cite{Tsai} that takes %into account effects of retardation. In this case Eliashberg's theory emerges as the fixed point. 

Going back to the result of the Fermi liquid fixed point, R. Shankar \cite{rs} discusses a $1/N$ expansion, with $N\simeq K_F/\Lambda$, $K_F$ the Fermi momentum which is finite, and $\Lambda \rightarrow 0$ as high-energy modes are eliminated, leading to $N \rightarrow \infty$. This large-$N$ analysis explains how this non-perturbative result can be obtained with the RG even when it is done only up to one-loop. Applications to related areas, including nuclear matter, quark matter, and quantum dots, are also briefly discussed. The emphasis of this article is to present the main ideas of the RG for interacting non-relativistic fermions to a wide audience. The readers are referred to the well-known review by the same author  \cite{RevModPhys.66.129} for further details.

Following the article on RG for non-relativistic fermions is Maria  Vozmediano's article \cite{mv} on RG studies for relativistic fermions that are present as the low-energy quasiparticles in graphene. Graphene is a one-atom-thick layer of carbon atoms covalently bonded forming a honeycomb lattice in two dimensions. Graphene has been the subject of extremely intense research since the ground-breaking experiments in 2004 by the Manchester group that led to the most recent Nobel Prize in Physics to Andre Geim  and Konstantin Novoselov.  Original references can be found in Ref. \cite{RevModPhys.81.109}. Vozmediano and her collaborators have studied electrons on a honeycomb lattice as early as 1994. Here, Vozmediano reviews some key results obtained in past and recent work. Following the same general strategy as described in R. Shankar's article \cite{rs}, the goal is to understand the low-energy physics of the problem. Unlike the case of non-relativistic fermions, however, for massless Dirac fermions in (2+1) dimensions, short-range interactions are irrelevant, but gauge interactions, such as the long-range Coulomb interaction, play an important role in the low-energy physics. It is pointed out that this system is different from QED in (2+1) dimensions because, while the electrons in graphene are confined in two spatial dimensions, the photons mediating the Coulomb interaction between the electrons live in three spatial dimensions. The dimensionless coupling constant $g = e^2/4\pi v_F$ in graphene plays the equivalent role of the fine structure constant in QED, with $v_F$, the Fermi velocity, replacing the speed of light $c$. Interestingly, it is found that the graphene system has a non-trivial infra-red fixed point where, instead of renormalizing to zero, the coupling constant flows to the value of the fine structure constant, $\alpha = 1/137$. While the charge of the electron does not renormalize, its velocity acquires logarithmic corrections, increasing until it becomes equal to $c$. Signs of marginal Fermi liquid behavior,  effects of disorder, and other aspects are also briefly discussed. Note that in recent years, the questions of chiral symmetry breaking and deconfinement, which are central in lattice gauge theory, 
have become increasingly important in graphene related work.  

Concepts being used in quantum information theory such as the entanglement entropy can provide new ways to 
look at the interplay between a subsystem and its environment. More generally, the way blocks 
of various sizes interact in quantum models is one of the basic question that  RG calculations try to answer. 
Ulrich Schollw\"{o}ck \cite{us} introduces the density-matrix renormalization group (DMRG) method. 
The DMRG has established itself
as a leading method for the simulation of the statics and
dynamics of one-dimensional strongly correlated quantum lattice systems. The 
DMRG is understood 
primarily as a variational method, but it can also be constructed as an RG flow in the space of 
reduced density operators. This dual nature is briefly discussed by Schollw\"{o}ck, and 
the DMRG method is presented in terms of Matrix Product States (MPS).  The construction relies on the singular value decomposition and admits nice graphical representations. 
The formulation provides simple expressions for the entanglement entropy and allows to assess the quality of approximations. 
The advantage for this approach is that the MPS states have a direct generalization to higher dimensions as Tensor Network States (TNS). Some generalizations of the DMRG to higher dimensions and the challenges involved are also discussed. For more details see \cite{Schollwšck201196,2010AnPhy.325.2153S}. 

\section{Applications in nuclear physics}

The renormalization group has had a major impact on atomic and nuclear physics. 
Renormalization group transformations must typically be approximated.  It is sometimes possible to  perturb about a free theory (i.e., a Gaussian fixed point) and accurately approximate ultraviolet scaling with a few relevant or marginal operators . The scaling classification of these operators is valid only in the region of the free theory, so when any couplings become unnaturally large we need a new approximation. When a non-Gaussian fixed point exists (i.e. a scale-covariant Hamiltonian or Lagrangian containing interactions, possibly strong), we can perturb about this fixed point  and use operators with good scaling behavior in the region of this fixed point. The number and nature of relevant, marginal and irrelevant operators can differ radically from those used in the region of a free theory. If a limit cycle exists, we must perturb about this cycle and once more our initial task is to find operators with good scaling behavior in the neighborhood of a limit cycle.
When applied to non-relativistic few-body physics, the RG reveals the existence of a non-Gaussian fixed point in the two-body system and a limit cycle in three-body systems. Cold atom and nuclear few-body systems display similar universal behavior that has been systematically classified and explored both theoretically and experimentally. 

Starting with Weinberg's seminal work on effective field theory (EFT) for low-energy nuclear physics in the early 90's, the use of EFT and RG methods \cite{Bogner:2009bt} has drastically improved our understanding of and computational control over nuclear two- and many-body interactions. Mike Birse  \cite{mb} briefly reviews chiral-EFT ($\chi$EFT) but his focus is a Wilsonian RG analysis of nuclear EFTs, which provides fixed points, limit cycles and operators with good scaling behavior that can be used to explore their neighborhoods. The two-nucleon contact interaction displays non-Gaussian fixed point behavior and Mike Birse maps out the region of this fixed point in detail. He then discusses how long-range two-nucleon interactions  (e.g. one-pion exchange) couple with short-range two-nucleon interactions to produce a rich array of scaling behavior for long-range three-nucleon interactions. The Efimov effect, a geometric tower of three-body bound states, is an example of what can result in non-relativistic few-body universality classes.

Hans Hammer and Lucas Platter \cite{lp} review Efimov physics and its four-body extension. Here again, the renormalization group reveals a wealth of universal few-body dynamics that has application in atomic and nuclear physics. They focus largely on cold atom systems, where few-body interactions can be tuned and one can directly explore RG trajectories in the regions of fixed points and limit cycles. In all of these physical systems the effective theory is valid up to a maximum cutoff beyond which couplings typically become unnaturally large. Physical hamiltonians do not lie on perfect RG trajectories but near them, and irrelevant operators must be identified and tuned to reproduce effective ranges, for example. Three-body universality is well-understood and recent progress on understanding four-body universality and its realization in cold atom systems is summarized.

There has been a lot of activity on energy-independent RG transformations applied to phenomenological and $\chi$EFT nuclear interactions. So-called $V_{low\ k}$ and Similarity Renormalization Group (SRG) approaches produce Hamiltonians in which far-off-shell interactions are suppressed, leading to softened nucleon-nucleon interactions and drastically improved convergence in variational calculations for nuclei. The ultimate goal of precision microscopic nuclear structure and reaction calculations has not been reached but with recent control of the running three-nucleon interaction and demonstration that the many-body problem becomes perturbative for sufficiently small cutoffs, we may be nearing this goal. These methods and a wide array of applications to nuclear systems has been recently reviewed in \cite{Bogner:2009bt}.

\section{Applications in particle physics}
As already seen in the discussion of the Efimov effect, scale invariance and conformal symmetry form a central theme in RG 
studies. 
Conformal symmetry 
can be destroyed or restored by tuning a parameter through some 
critical value. The appearance of a mass gap can break the conformal symmetry of the microscopic theory. 
The mechanism of dimensional transmutation in asymptotically free theories allows to express a physical mass (gap) $m$ approximately as  
\begin{equation}
ma \propto {\rm e}^{-A/g^2(a)} \  ,
\end{equation}
with $a$ is the lattice spacing or the inverse of an ultraviolet regulator, $A$ a positive constant obtained from a one-loop calculation and $g(a)$ the running coupling constant. 
Asymptotic freedom means that in the continuum limit, where $a$ goes to zero, $g^2(a)$ becomes small in such a way that the physical value $m$ stays constant.
On the other hand, at large distance one expects confinement: it is impossible to 
liberate the quarks inside protons and neutrons at least at zero temperature.  

The question of confinement is reviewed from a RG point of view in Mike Ogilvie' s article \cite{mo} who introduces basic concepts such as Wilson loops, Polyakov lines and string tension. 
Confinement can be seen as a global property of some RG flows. 
The linearized behavior of RG flows near a fixed point has universal features that can be expressed in terms of critical exponents. 
However, calculating the RG flows between fixed points is usually a difficult 
nonlinear problem. QCD confinement can be stated as the smoothness of RG flows 
between the weakly interacting fixed point where asymptotic freedom holds and a strongly coupled fixed point where a linear potential and a mass gap are manifest.  This is the outline of a possible proof of confinement for pure gauge theory \cite{Tomboulis:2009zz}.
More elaborate schemes are needed to elucidate the modifications to the 
IR structure due to the presence of a sufficient number of fermions or Polyakov line terms. Recent progress about this question based on semi-classical methods, Nambu-Jona Lasinio models, functional RG methods and Schwinger-Dyson equations are reviewed in Mike Ogilvie's article \cite{mo}. 

In particle physics, models based on mechanisms having the same effects as the so-called Higgs mechanism but without fundamental scalars fields have been proposed over the last three decades. 
References for some of these models beyond the standard model can be found in Ref. \cite{Sannino:2009za}. Phenomenological constraints require a slow running coupling constant. This situation is in some sense 
 ``nearly conformal".  
Tom deGrand's article \cite{td} reviews recent simulations in lattice gauge theory which provide indications of new infrared (IR) fixed points in models similar to  the standard theory of strong interactions (Quantum ChromoDynamics or QCD for short) but with more flavors or higher order representations. Exactly at these IR fixed point, there is no special scale and conformal symmetry is manifest.  

The running of the coupling constant has been calculated with different methods. 
Graphs of RG flows or running of couplings look quite similar to those encountered in Hubbard models with competing next-to-nearest-neighbor interactions or Bose-Hubbard models, calculated with functional RG methods. 
For some range of values of the number of flavors where asymptotic freedom still holds (called the ``conformal window"), a nontrivial IR fixed point can be found. 
The continuum physics of these gauge theories is quite different from 
QCD: conformal symmetry instead of confinement and chiral symmetry 
breaking. New observables are needed for this situation, and the term
``unparticle physics" has been coined for it. Recent progress in determining the conformal windows for various type of lattice gauge theories are reviewed in deGrand's article \cite{td}.

\section{Applications of the ERG}

The idea of exact RG (ERG) equations was proposed more than thirty years ago \cite{wilson74,wegner76}. 
In section 11 of \cite{wilson74}, the terminology ``exact renormalization group equations in differential form" was used and the abbreviation ERG
is a generic way to refer to sets of differential equations that control the running of all the possible couplings. 
This idea was reconsidered very actively in the early nineties. This led to many different applications reviewed in Ref. \cite{berges00}. 
This has become a very active and broad field of investigation. Conferences on this topic are held every two years. The most recent, organized at the Corfu Summer Institute had more than one hundred participants. In general, ERG equations require approximations in order to be solved. These approximations usually introduce a dependence on the choice of the cutoff function defined below. For this reason, the word ``exact''  in ERG is sometimes replaced by ``functional''. 
  
The most common  ERG equations involve the addition of a quadratic term 
\begin{equation}
\Delta S_k[\phi]=\int d^d q\phi_AR^{AB}_k(q^2)\phi^B \  ,
\end{equation}
to the microscopic (``bare") action. 
This term depends on the  cutoff function $R^{AB}_k(q^2)$ which needs to be chosen in order to suppress the modes $\phi_A(q)$ such that 
$q<k$. For this reason,  $k$ is often called the infrared cutoff. The change in the effective action as $k$ is lowered can be calculated using 
a simple Schwinger-Dyson equation.  As $k\rightarrow0$, we recover the effective action if a suitable cutoff function is chosen. 
Discussion of various types of equations (the so-called Polchinski's equation or Wetterich' s equation) and their relationship are discussed in detail in \cite{bagnuls00,2007cond.mat..2365D,Igarashi:2009tj}. There are also review articles or lecture notes that focus on gauge theories \cite{Becchi:1996an,Gies:2006wv,Pawlowski:2005xe} or quantum gravity \cite{Percacci:2007sz,Litim:2008tt}.

One important test of a new RG method is the calculation of the critical exponents  for which very accurate estimates are often available
\cite{ZinnJustin:2002ru,Pelissetto:2000ek}.
Jean-Paul Blaizot \cite{jpb} presents the ``BMW" approximation scheme for the ERG. 
After reviewing the basic ERG equations discussed above, he suggests an economical way to take into account the momentum dependence of the $n$-point functions. 
This goes beyond the local-potential approximation and  with
a regulator satisfying the principle of minimal sensitivity, a numerical 
treatment can provide excellent results for critical exponents and other
quantities at a modest cost. The applications include the critical exponents for $O(N)$ models and the shift of the transition temperature in Bose-Einstein condensation of dilute gases. 

Quantum gravity has been a theoretical puzzle for many years. The non-renormalizabilty of the 
conventional Feynman diagram approach has generated many interesting alternative formulations. 
Weinberg's scenario of ``asymptotic safety" is the idea that quantum gravity 
might be controlled by a nontrivial UV fixed point, so that well-defined 
predictions can be made using a gravitation field theory without 
invoking any underlying, more microscopic degrees of freedom. Daniel Litim \cite{dl}
reviews  recent progress in this area, based on the ERG. A UV fixed point has 
been found for the Einstein action with a cosmological constant, and has 
been shown to be stable against addition of polynomials in the Ricci 
scalar and the square of the Weyl tensor. The reconsideration of the one-loop calculations in this context 
emphasizes the role of quadratic divergences which are absent in dimensional regularization. The effects of 
small extra dimensions on the RG flows are also discussed.

Michael Scherer, Stefan Floerchinger, and Holger Gies review \cite{hg} work on the BCS-BEC crossover using the ERG method. 
In a quantum gas of fermionic atoms with different hyperfine states, the interaction strength can be tuned with an external magnetic field. As the magnetic field is increased and tuned across a Feshbach resonance, the effective scattering length goes from positive to negative values, diverging with opposite sign as the resonance is approached from either side. In the strong interaction regime, perturbative approaches break down. Scherer {\it et al.} present a non-perturbative treatment for the universal features at unitarity as well as the BEC and BCS limits, all captured within the same truncation scheme. The question of regulator-scheme dependence that arises in ERG treatments is addressed by comparison with well-known results in the weak-coupling limits. The authors calculate the phase diagram and compare their results with previous analytical and numerical studies. 

\section{Topics not covered}
In this theme issue, we have unfortunately left out reviews of the RG approach to solutions of partial differential equations, with applications to fluid dynamics and cosmology \cite{goldenfeld07} and also RG applications in the 
AdS/CFT correspondence context
 \cite{Heemskerk:2010hk}.
Some new application of the RG methods have also emerged, for instance in optical lattices \cite{DBLP:reference/complexity/MatheyTN09}, studies of conformal symmetry and confinement in terms of complex RG flows
 \cite{Kaplan:2009kr,moroz09,Denbleyker:2010sv,onprogress,hmprogress}, studies of the relationship between discrete and continuous RG flows
 \cite{hmreview,discprogress,Curtright:2010hq} .
\section{Conclusions and acknowlegments}

In summary, the RG method is an important tool to deal with many interesting but difficult problems in physics. 
It has become obvious that communication across traditional field boundaries is very profitable and that a RG community 
is building up. We hope that  this theme issue will contribute in this process. 

%\begin{acknowledgments}
The idea of the theme issue started  during the workshop 
``New applications of the renormalization group
method in nuclear, particle and condensed matter physics" held at the Institute for Nuclear Theory, University of Washington, Seattle (INT-10-45W). 
We thank David Kaplan and the INT staff for their support and Mike Birse for co-organizing the workshop. 
Additional work was done at the Aspen Center for Physics in May and June 2010 during the workshop ``Critical Behavior of Lattice models". 
We thank the participants of these two workshops 
for stimulating discussions. 

This 
research was supported in part  by the Department of Energy
under Contract No. FG02-91ER40664, the 
National Science Foundation
under grant DMR-0847801 and UC-Lab FRP under award number
09-LR-05-118602.
%\end{acknowledgments}
%\bibliographystyle{aps}
%\bibliographystyle{abbrv}
%\bibliographystyle{phjcp}
%\bibliographystyle{unsrt}
%\bibliographystyle{ieeetr}
%\bibliography{macbib}
%\input{INTroARTF}

\end{document}